\documentclass[onecolumn,draftcls]{IEEEtran}
\ifCLASSINFOpdf
\else
\fi
\usepackage[normalem]{ulem}
\usepackage{cite}
\usepackage{graphicx}
\usepackage{hyperref}
\usepackage{amsmath}
\usepackage{amssymb}
\usepackage{bm}
\usepackage{algorithmic}
\usepackage{algorithm}
\usepackage{array}
\usepackage{booktabs}
\usepackage{multirow}
\usepackage{makecell}
\newcommand \F{\mathbb{F}}
\newcommand \X{\bm{x}}
\newcommand \Y{\bm{y}}

\newcommand \e{\bm{e}}
\newcommand \E{\bm{E}}
\newcommand \C{\mathcal{C}}

\newcommand \V{\bm{v}}
\newcommand \U{\bm{u}}
\newcommand \sig{\bm{\sigma}}
\newcommand \T{\mathrm{T}}
\newcommand \zero{\underline{0}}
\DeclareMathOperator{\tail}{tail}
\DeclareMathOperator{\head}{head}
\DeclareMathOperator{\Out}{Out}
\DeclareMathOperator{\In}{In}

\DeclareMathOperator{\ddeg}{deg}

\newtheorem{defi}{Definition}
\newtheorem{exam}{Example}
\newtheorem{prop}{Proposition}
\newtheorem{rema}{Remark}

\begin{document}
%
\title{Distributed Decoding of Convolutional Network Error Correction Codes}
%
%
%

\author{Hengjie~Yang
        and~Wangmei~Guo
\thanks{This paper was submitted in part at the 2017 IEEE International Symposium on Information Theory, Aachen, Germany.}
\thanks{Hengjie Yang is with the School of Telecommunications Engineering, Xidian University, Xi'an, China (e-mail: yanghengjieyhj@gmail.com)}
\thanks{Wangmei Guo is with the State Key Lab of Integrated Services Networks, Xidian University, Xi'an, China (e-mail: wangmeiguo@xidian.edu.cn)}}

\maketitle

\begin{abstract}
A Viterbi-like decoding algorithm is proposed in this paper for generalized convolutional network error correction coding. Different from classical Viterbi algorithm, our decoding algorithm is based on minimum error weight rather than the shortest Hamming distance between received and sent sequences. Network errors may disperse or neutralize due to network transmission and convolutional network coding. Therefore, classical decoding algorithm cannot be employed any more. Source decoding was proposed by multiplying the inverse of network transmission matrix, where the inverse is hard to compute. Starting from the \textit{Maximum A Posteriori (MAP)} decoding criterion, we find that it is equivalent to the minimum error weight under our model. Inspired by Viterbi algorithm, we propose a Viterbi-like decoding algorithm based on minimum error weight of combined error vectors, which can be carried out directly at sink nodes and can correct any network errors within the capability of convolutional network error correction codes (CNECC). Under certain situations, the proposed algorithm can realize the distributed decoding of CNECC.
\end{abstract}


%
\IEEEpeerreviewmaketitle

\section{Introduction}
%
%
%
%
\IEEEPARstart{N}{etwork} coding is a new technique introduced in \cite{ACLY2000} which allows nodes to make the combination of multiple information before forwarding it. It is shown with large advantages in throughput, load equalization and security and so on, and has attracted lots of attention \cite{005}\cite{KM2003}. Network error correction coding was first proposed by Cai \& Yeung to correct errors caused by adversaries, which was then completely introduced in\cite{YC2006}\cite{YC20062}. They extended the Hamming bound, Singleton bound and Gilbert-Varshamov bound from classical error correction coding to network coding. Refined coding bounds for network error correction were given in \cite{YY2007}. Zhang studied network error correction in packet networks \cite{001}, where an algebraic definition of the minimum distance for linear network codes was introduced and the decoding problem was studied. Network error detection by random network coding has been studied by Ho \textit{et al.} \cite{HLKMEK2004}. Jaggi \textit{et al.} \cite{JLKHKME2008} have developed random algorithms for network error correction with various assumptions on adversaries. A new general framework for non-coherent network error correction was introduced in \cite{KK2008}. In their framework, messages are modulated as subspaces, so a code for non-coherent network error correction is also called a subspace code. Using rank-metric codes, nearly optimal subspace codes are constructed and decoding algorithms are also studied in \cite{SKK2008}.

Convolutional network coding is shown to have advantages in field size, small decoding delay and so on, which is more suitable for practical communications \cite{LSZ2011}\cite{GSCM2013}. Convolutional network-error correcting coding was introduced in \cite{004} in the context of coherent network coding for acyclic instantaneous or unit-delay networks. They presented a convolutional code construction for a given acyclic instantaneous or unit-delay memory-free network that corrects a given pattern of network errors. For the same network, if the network code changes, then the convolutional code obtained through their scheme may also change. They also consider the decoding. Decoding may be carried out at source or sink nodes based on the distance of equivalent received sequences. If the decoding needs to be done at the source node, the inverse of network transmission matrix is multiplied to the received sequences at a sink node, and then messages are decoded with the classical Viterbi algorithm. Distance measures for convolutional codes in rank metric were given in \cite{002}, and two constructions of (partial) unit memory ((P)UM) codes in rank metric based on the generator matrices of maximum rank distance codes were presented to correct network errors for non-coherent multi-shot networks. They also provided an efficient decoding algorithm based on rank-metric block decoders for error correction in random linear network coding.

In this paper, we consider the decoding for a generalized convolutional network error-correction code constructed by the extended method in \cite{004}, where network errors on each edge occur with the same probability and are separated by fixed timeslots. Source encoding is always needed due to the error correction purpose in networks. 

However, given a delay-invariant, single source multicast network with a generalized multicast convolutional code, the ideal scenario is to decode CNECC in a distributed way while ensuring the decoded sequence is identical to the input sequence, which we refer to as the distributed decoding of CNECC. To realize it, we propose a Viterbi-like decoding algorithm based on minimum error weight and give a sufficient condition to show the feasibility of such decoding process.

The main contributions of this paper are as follows:
\begin{itemize}
  \item A Viterbi-like decoding algorithm is proposed based on minimum error weight for generalized convolutional error correction coding.
  \item The algorithm can be carried out directly at sink nodes and no additional processing matrix is required.
	\item A sufficient condition of realizing the distributed decoding of CNECC is first given.
\end{itemize}

This paper is structured as follows. In Section \ref{sec2}, definitions and notation for generalized convolutional network error correction coding as well as the relation between error and distance are given. Section \ref{secI} gives a sufficient condition of realizing the distributed decoding of CNECC. Section \ref{sec3} illustrates our Viterbi-like decoding algorithm and performance analysis. Some examples are shown in Section \ref{sec4} and simulation results are given in Section\ref{secII}. Finally, Section \ref{sec5} concludes this paper.

\section{Definitions and Notation}\label{sec2}
\subsection{Network model}
In this paper, we consider the finite directed network as in \cite{001}. A finite directed graph $G$ can be represented as $\{V,E\}$ where $V$ is the set of all vertices in the network and $E$ is the set of all edges in the network. A directed edge $e=(i,j)$ represents a channel leading from node $i$ to node $j$ where $i$ is called the tail of $e$ and $j$ is called the head of $e$, i.e., $\tail(e)=i$ and $\head(e)=j$. Channel $e$ is called an outgoing channel for node $i$ and an incoming channel for node $j$. For a node $i$, let $\In(i)=\{e\in E:\text{$e$ is an incoming channel of $i$}\}$ and $\Out(i)=\{e\in E:\text{$e$ is an outgoing channel of $i$}\}$.

Let $S$ and $T$ be two disjoint sets of $V$. The elements in $S$ are called source nodes which only have outgoing channels. The elements in $T$ are called sink nodes which only have incoming channels. The rest of nodes in set $I=V-S-T$ are called internal nodes. In this paper, we only consider single source networks and denote by $s$ the unique source node.

We assume that each edge in the network has unit capacity (can carry utmost one symbol from $\F_q$) and capacity between nodes greater than one is modeled as parallel edges. A cut between node $i$ and node $j$ is a set of edges whose removal will disconnect $i$ and $j$. For unit capacity channels, the capacity of a cut is regarded as the number of edges in it. For a sink node $t\in T$, let $\omega_t$ be the unicast capacity between $s$ and $t$. Then, $\omega=\min_{t\in T}\omega_t$ is the max-flow min-cut capacity of the multicast connection. Here we let the information rate be $\omega$ symbols per time instant and we only consider multicast networks.

\subsection{Network codes}
We follow \cite{KM2003} in describing network codes. An $\omega$-dimensional network code can be described by three matrices (over $\F_q$), $A_{\omega\times|E|},\ K_{|E|\times|E|}(z),\ B^t_{|E|\times\omega}$ (for every sink node $t\in T$). The details of them can be found in \cite{KM2003}.
\begin{defi}[\cite{KM2003}]
The network transfer matrix, $M_t(z)$, corresponding to a sink node $t\in T$ for an $\omega$-dimensional network code, is a full-rank (over the field of rationals $\F_q(z)$) $\omega\times\omega$ matrix defined as
\[
	M_t(z):=A(I_{|E|\times|E|}-K(z))^{-1}B^t=AF_t(z)
\]
\end{defi}
In this paper, we study network codes over $\F_2$.

\subsection{CNECC}
We follow \cite{004} in describing a CNECC. Assume the $\omega$-dimensional network codes have been implemented in the given single source network. Note that the network may be a time-delay network, i.e. the network whose matrix $K$ carries delay factor $z$. The following definitions describe a convolutional code for error correction in such networks.
\begin{defi}[\cite{004}]
An input convolutional code, $\C_s$, is a convolutional code of rate $k/\omega\ (k<\omega)$ with an input generator matrix $G_I(z)$ implemented at the source of the network.
\end{defi}
\begin{defi}
The output convolutional code, $\C_t$, corresponding to a sink node $t\in T$, is the $k/\omega\ (k<\omega)$ convolutional code generated by matrix $G_{O,t}(z)$ which is given as $G_{O,t}(z)=G_I(z)M_t(z)$, with $M_t(z)$ being the full-rank network transfer matrix corresponding to an $\omega$-dimensional network code.
\end{defi}
\begin{defi}
The free distance of the convolutional code $\C$ is given as
\[
	d_{free}(\C)=\min\{w_H(\V(z))|\V(z)=\U(z)G(z)\in\C,\V(z)\ne 0\}
\]
where $w_H$ indicates the Hamming weight over $\F_q$.
\end{defi}
\begin{defi}[\cite{004}]\label{defi05}
Let $\C$ be a rate $b/c$ convolutional code with a generator matrix $G(z)$. Then corresponding to the information sequence $\U_0,\U_1,\dots(\U_i\in\F^b_q)$ and the codeword sequence $\V_0,\V_1,\dots(\V_i\in\F^c_q)$, we can associate an encoder state sequence $\sig_0,\sig_1,\dots,$ where $\sig_i$ indicates the content of the delay elements in the encoder at a time instant $i$. Define the set of $j$ output symbols as $\V_{[0,j)}:=[\V_0,\V_1,\dots,\V_{j-1}]$ and the set $S_{d_{free}}$ as follows.
\[
	S_{d_{free}}:=\{\V_{[0,j)}|w_H(\V_{[0,j)})<d_{free}(\C),\sig_0=\bm{0},\forall j>0\}
\]
Clearly, the definition of $S_{d_{free}}$ excludes the possibility of a zero state in between, i.e., $\sig_i\ne0$ for any $0<i\le j$. We have that the set $S_{d_{free}}$ is invariant among the set of minimal convolutional encoders. We now define
\[
	T_{d_{free}}(\C):=\max_{\V_{[0,j)}\in S_{d_{free}}}j+1
\]
which thereby can be considered as a code property because of the fact that $S_{d_{free}}$ is invariant among minimal encoders.
\end{defi}
\begin{defi}
An error pattern $\rho$ is a subset of $E$ which indicates the edges of the network in error. An error vector $\e$ is a $1\times |E|$ vector which indicates the error occurred at each edge. An error vector $\e$ is said to match an error pattern if all nonzero components of $\e$ occur only on the edges in $\rho$.  
\end{defi}

Let $\X(z),\ \Y(z),\ \e(z)$ be the input sequence, output sequence and network error sequence, respectively. Thus, at any particular sink node $t\in T$, we have
\begin{align}
	\Y(z) &=\X(z)G_I(z)M_t(z)+\e(z)F_t(z)\notag\\
			&=\X(z)G_{O,t}(z)+\e(z)F_t(z)\label{eq1}
\end{align}

In Section IV \cite{004}, the authors proposed a construction scheme that can correct a given set of error patterns as long as consecutive network errors are separated by a certain interval. The convolutional code constructed under their scheme is a CNECC. They also proposed two cases of decoding such CNECC under different conditions.

\section{Distributed Decoding of CNECC}\label{secI}
We now study the characteristics of matrix $F_t(z)$ at sink node $t\in T$. Assume the degree of $F_t$ is $l_t=\ddeg(F_t(z))$. Thus, $F_t(z)$ can be written as $F_t(z)=\sum_{i=0}^{l_t}F_iz^i$ where $F_{l_t}$ is a nonzero matrix. For a particular error vector $\e$, we can get its resulting combined error vector $(\e F_0\ \e F_1\ \cdots\ \e F_{l_t})$ where $\e F_i\ (0\le i\le l_t)$ is a $1\times\omega$ subvector from \eqref{eq1}. Combined error vector characterizes the impact of an error vector $\e$ in the network. If all combined error vectors can be corrected, then all network errors can be consequently corrected. Let $\bm{E}'_i=(\e_i F_0\ \e_i F_1\ \cdots\ \e_i F_{l_t})$ be the combined error vector generated by error vector $\e_i$ in which only the $i^{th}$ bit is one and the rest are all zeros. Let $L$ be a collection of vectors in a linear space and $\langle L\rangle$ represent the subspace spanned by the vectors in $L$. Here we consider the subspace $\Delta(t,l_t)=\langle\{\bm{E}'_i:1\le i\le E\}\rangle$ at sink node $t\in T$.

Note that $\Delta(t,l)$ can also be regarded as a subspace determined by parameter $l$ where $l\ge l_t$. For a given set of error patterns, assume $G_I(z)$ has been constructed using the scheme in \cite{004}. We hope to find a minimum $l$ such that $\Phi(t,l)\cap\Delta(t,l)=\{\zero\}$ where $\zero$ is a $1\times \omega(l+1)$ zero vector and $\Phi(t,l)$ is the message subspace spanned by output convolutional codes generated by $\X_l(z)G_{O,t}(z)$ where $\X_l(z)$ are all possible input sequences from instant $0$ to $l$, i.e., $(00\cdots00)_l,\ (00\cdots01)_l,\ \cdots,\ (11\cdots11)_l$.

Before discussing how to realize the distributed decoding of a CNECC, we give the following proposition.
\begin{prop}\label{prop1}
Assume $l$ is available at sink node $t\in T$ such that  $\Phi(t,l)\cap\Delta(t,l)=\{\zero\}$. Then in any sliding window with length $l+1$ on the output trellis of $G_{O,t}(z)$, at most one nonzero combined error vector exists as long as network errors are separated by $l+1$ timeslots.
\end{prop}
\begin{IEEEproof}
We use reduction to absurdity to prove this proposition. Before the proof, we assume all sliding window mentioned below share the same length $l+1$. Apparently, proving all situations in first window hold true is enough as all situations in succeeding windows are equivalent to that in first window by removing the impact of the input sequence prior to current window. Assume in first window we find two different nonzero combined error vectors $\e_1,\e_2 \in \Delta(t,l)$ which implies that there exists two different input sequences $\X_1,\X_2$ satisfying $f(\X_1)+\e_1=f(\X_2)+\e_2$ where $f(\X)=\X(z)G_{O,t}(z)\in \Phi(t,l)$. Given the closure property of $\Phi(t,l)$ and $\Delta(t,l)$, i.e., $f(\X_1)-f(\X_2)=f(\X_1-\X_2)=f(\X_3)\in\Phi(t,l)$ and $\e_2-\e_1=\e_3\in\Delta(t,l)$ where $\X_3,\e_3$ are another input sequences and nonzero combined error vector in first window respectively, it leads to $f(\X_3)=\e_3$ which contradicts to the condition $\Phi(t,l)\cap\Delta(t,l)=\{\zero\}$. The proof is completed.
\end{IEEEproof}

Proposition \ref{prop1} indicates the uniqueness of nonzero combined error vector within each sliding window when network errors are separated by a certain interval. However, the crux of realizing the distributed decoding of CNECC is how to pinpoint the unique time instant when network errors occur. That is, the addition of two combined error vectors with one generating at current instant and another generating at a later instant may yield a certain correct output sequence. Under the circumstances, we cannot distinguish exactly when network errors occur as both time instants are possible. In fact, if an addition yields a correct output sequence with $\V_0=\bm{0}$, we can subjectively determine the network errors occur at the time instant of which is earlier than another. We call it error preposing which can enable the decoding algorithm still to proceed. The real issue is how to determine network errors when an addition yields a correct output sequence with $\V_0\ne\bm{0}$. The following proposition gives a sufficient condition that is able to realize the distributed decoding of CNECC.
\begin{prop}\label{prop02}
The distributed decoding of CNECC at sink node $t\in T$ can be realized when $G_{O,t}(z)$ is non-catastrophic and its free distance $d_{free}\ge 2\omega(l_t+1)+1$.
\end{prop}
\begin{IEEEproof}
The main feature of non-catastrophic encoder is that only finite-length consecutive zero outputs exist. Similar to Definition \ref{defi05}, define $W(l)$ as follows.
\[
	W(l):=\min\{w_H(\V_{[0,l]})|l\ge0,\sig_0=\bm{0}\}
\]
where $\V_{[0,l]}:=\{\V_0,\V_1,\dots,\V_l\}$. Given the fact that $G_{O,t}(z)$ is non-catastrophic, $W(l)$ is an increasing function as $l$ increases, though at some time instant $W(l)$ may stay the same temporarily. For $l\ge T_{d_{free}}$, $W(l)=d_{free}$. Given the fact that the addition of arbitrarily two combined error vectors (meaning that they can generate at different time instants) can affect at most $2\omega(l_t+1)$ bits, there must exist a threshold $l_{gate}$ such that $W(l_{gate})\ge2\omega(l_t+1)+1$. Let $l=l_{gate}$, therefore in the first sliding window $[0,l]$, none of the addition of arbitrarily two combined error vectors is identical to a certain correct output sequence, indicating that the exact time instant when network errors occur can be pinpointed. By using mathematical induction, network errors can be pinpointed in all subsequent windows which implies that the distributed decoding is realized. The proof is completed.
\end{IEEEproof}

Proposition \ref{prop02} gives a sufficient condition that can realize the distributed decoding of CNECC. In fact, introducing $l_{gate}$ is to prove the feasibility of such decoding process. The real threshold value may be less than $l_{gate}$ as the fundamental condition of realizing the distributed decoding of CNECC is that no addition of arbitrarily two combined error vectors can yield a certain correct output sequence with $\V_0\ne\bm{0}$ in the first sliding window. $l_{gate}$ is merely a sufficient value that is able to meet this condition. An example is shown in Section \ref{sec4} to clarify this point.

\pagebreak
\section{Decoding Algorithm of CNECC}\label{sec3} 
Assume $l$ has been determined satisfying $\Phi(t,l)\cap\Delta(t,l)=\{\zero\}$ and we consider the following model.
\begin{enumerate}
\item all single edge errors have the same error probability $p$ and network errors are subject to i.i.d. under this distribution.
\item all network errors are separated by $l+1$ timeslots and only occur at prior $|\X(z)|$ time instants.
\item $G_{O,t}(z),F_{t}(z)$ are available at each sink node.
\end{enumerate}
Clearly, MAP is equivalent to the minimum error weight in this model so we propose the following decoding algorithm that is able to find the MAP path based on this model.
\begin{algorithm}[H]
\caption{Decoding algorithm at sink node $t\in T$}
\begin{algorithmic}[1]
\STATE Make reference table $\Delta(t,l)$. For each combined error vector, its weight is defined as that of its corresponding minimum error vector.
\WHILE{new $\Y_i\ (0\le i\le |\Y(z)|-1)$ is received}
\IF{$i\ge l$}
\STATE Calculate new combined error vectors $(\E_0,\dots,\E_l)$ from previous vectors with finite weight.
\IF{multiple $(\E_0,\dots,\E_l)$s converge to the same output of the same next state}
\STATE Select the one with minimum weight as father.
\ENDIF
\STATE $N\leftarrow$ the number of vectors in current window
\FOR{$j=0$ to $N-1$}
\STATE $(\E_0,\dots,\E_l)$ is the $j^{th}$ combined error vector.
\IF{$\E_0=\bm{0}$}
\STATE Set its accumulative weight as its father's.
\ELSIF{$(\E_0,\dots,\E_l)\notin\Delta(t,l)$}
\STATE Set its accumulative weight infinite.
\ELSE
\STATE Calculate its accumulative weight.
\STATE Remove it on output trellis.
\ENDIF
\ENDFOR
\IF{The input can be uniquely determined}
\STATE Output the newly found input sequence.
\ENDIF
\STATE Slide current window one instant forward
\ENDIF
\ENDWHILE
\end{algorithmic}
\end{algorithm}
\begin{rema}
In the above pseudo-code, $|\Y(z)|=|\X(z)|+\max\{\ddeg(G_{O,t}(z)),l\}$ since we need to encompass all combined error vectors completely. Compared to the decoding algorithm in \cite{004}, our algorithm has the following advantages. First, it can work directly at each sink node and no additional processing matrix has to be multiplied. Second, our algorithm can find the MAP path regardless of the error correction capability of $G_{O,t}(z)$ as in networks, MAP is equivalent to the minimum error path under our model. Third, the performance of the proposed algorithm is closely related to the characteristics of $F_t(z)$ and the free distance of $G_{O,t}(z)$. If conditions in Proposition \ref{prop02} can be met at each sink node $t\in T$, the distributed decoding process can be realized on the whole network when all network errors are separated by $l_{\max}+1$ timeslots where $l_{\max}$ is the maximum of thresholds of all sink nodes.

In fact, the whole decoding process is quite similar to the classical Viterbi algorithm as we decode messages purely based on the metric of minimum error weight. Under our model, the impact of an error vector, i.e., its corresponding combined error vector, has no overlap with one another on output trellis so that the algorithm could determine its corresponding minimum error vector by searching the reference table.
\end{rema}

Next, we study the time complexity of the proposed algorithm. Assume the reference table $\Delta(t,l)$ is available at each sink node $t\in T$ and at any particular time instant, the maximum processing time, including searching reference table, calculating Hamming weight, etc., is $C$. The total time complexity is $O(Cn)$.

\section{Illustrative Examples}\label{sec4}
\begin{exam}\label{exam01}
To illustrate the concept introduced in above sections and to show the complete decoding process of the proposed algorithm, we check a simple directed cyclic network shown in Fig. \ref{fig1}.
\begin{figure}[h]
\centering
\includegraphics[scale=0.8]{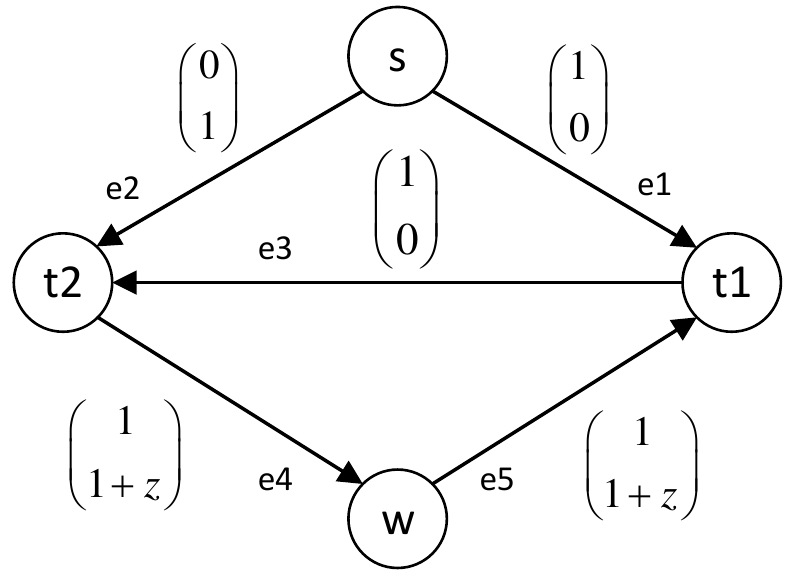}
\caption{Network $G_1$, with $\omega=2$}
\label{fig1}
\end{figure}
With the given network code, we thus have the network transfer matrices at sink $t_1$ and $t_2$ as follows
\[
	M_{t_1}(z)=\begin{bmatrix}1&1\\0&1+z \end{bmatrix}=AF_{t_1}(z)
\]
where
\[
	F_{t_1}(z)=\begin{bmatrix}1 & 0 & 0 & 0 & 0\\1 & 1+z & 1 & 1 & 1\end{bmatrix}^\T
\]
and
\[
	M_{t_2}(z)=\begin{bmatrix}1&0\\0&1 \end{bmatrix}=AF_{t_2}(z)
\]
where
\[
	F_{t_2}(z)=\begin{bmatrix}1 & 0 & 1 & 0 & 0\\0 & 1 & 0 & 0 & 0\end{bmatrix}^\T
\]

As $M_{t_2}(z)$ is an identity matrix, therefore $G_{O,t_2}(z)=G_I(z)$ at sink node $t_2$. Thus, we only consider sink node $t_1$ in later discussion. By using the construction scheme in \cite{004}, we choose $G_I(z)=[1+z^2\ 1+z+z^2]$ with $d_{free}=5$. Thus, $G_{O,t_1}(z)=G_I(z)M_{t_1}(z)=[1+z^2\ z^2+z^3]$ with $d_{free}=4$. 

Next, we illustrate $l=2$ satisfies the condition that $\Phi(t_1,l)\cap\Delta(t_1,l)=\{\zero\}$. Let $F_{t_1}(z)=F_0+F_1z+\bm{0}z^2$ where $\bm{0}$ is an $|E|\times\omega$ zero matrix. Thus, $\Phi(t_1,1)$ and $\Delta(t_1,l)$ are given in Table \ref{table1} and \ref{table2}. Elements of $\Phi(t_1,l)$ with $l=2$ are listed in column 2 of Table \ref{table1} and elements of $\Delta(t_1,l)$ with $l=2$ are listed in column 3 of Table \ref{table2}. Thus, it can be easily checked that only $\zero$ is in the intersection of $\Phi(t_1,l)$ and $\Delta(t_1,l)$.
\begin{table}[t]
\centering
\caption{The subspace of $\Phi(t_1,l)$ with $l=2$}
\begin{tabular}{cc}
\toprule
 $\X(z)$ in window $[0,l]$ & $\X(z)G_{O,t_1}(z)$ in window $[0,l]$  \\\midrule
000 & 00 00 00 \\
001 & 00 00 10 \\
010 & 00 10 00 \\
011 & 00 10 10 \\
100 & 10 00 11 \\
101 & 10 00 01 \\
110 & 10 10 11 \\
111 & 10 10 01 \\
\bottomrule
\end{tabular}
\label{table1}
\end{table}

\begin{table}[t]
\centering
\caption{The subspace of $\Delta(t_1,l)$ with $l=2$ (reference table)}
\begin{tabular}{cccc}\toprule
index & minimum error vector $\e$ & $\e F_0\ \e F_1\ \e\bm{0}$  & weight \\\midrule
0 & $(00000)$ & 00 00 00 & 0\\\midrule
1 & $(10000)$ & 11 00 00 & 1\\\midrule
2 & $(01000)$ & 01 01 00 & 1\\\midrule
3 & $(00100)$ $(00010)$ $(00001)$ & 01 00 00 & 1\\\midrule
4 & $(11000)$ & 10 01 00 & 2\\\midrule
5 & $(10100)$ $(10010)$ $(10001)$ & 10 00 00 & 2\\\midrule
6 & $(01100)$ $(01010)$ $(01001)$ & 00 01 00 & 2\\\midrule
7 & $(11100)$ $(11010)$ $(11001)$ & 11 01 00 & 3\\
\bottomrule
\end{tabular}
\label{table2}
\end{table}

Assume the input sequence $\X(z)=1+z^2+z^5$ with $|\X(z)|=6$. We set two types of $\e(z)$ where the fomer type contains indistinguishable network errors and the latter type is the opposite. They are given as follows.
\begin{align*}
	\e_{\alpha}(z)&=(11000)+(00000)z+(00000)z^2+(10100)z^3+(00000)z^4+(00000)z^5\\
			&=(1+z^3\ 1\ z^3\ 0\ 0)\\
	\e_{\beta}(z)&=(10000)+(00000)z+(00000)z^2+(00100)z^3+(00000)z^4+(00000)z^5\\
			&=(1\ 0\ z^3\ 0\ 0)
\end{align*}
With equation \eqref{eq1}, we have the following output sequences.
\begin{align*}
	\Y_{\alpha}(z)&=\X(z)G_{O,t_1}(z)+\e_{\alpha}(z)F_{t_1}(z)\\
					&=(00)+(01)z+(01)z^2+(11)z^3+(11)z^4+(11)z^5+(00)z^6+(11)z^7+(01)z^8\\
	\Y_{\beta}(z)&=\X(z)G_{O,t_1}(z)+\e_{\beta}(z)F_{t_1}(z)\\
					&=(01)+(00)z+(01)z^2+(00)z^3+(11)z^4+(11)z^5+(00)z^6+(11)z^7+(01)z^8
\end{align*}

Now, we illustrate the decoding process when exerting network error sequence. At first, we list all combined error vectors (in decimal form) between $\Y(z)$ and the output of trellis within the sliding window. The accumulative weight of each combined error vector at current sliding window is given in parentheses. Those with infinite weight are marked with a strikeout which indicate they cease to extend. At the next time instant, new combined error vectors are derived from previous extendable states by removing the impact of its combined error vector and appending $\omega$ bits of new combined errors. At each time instant, all combined error vectors are listed in an increasing order of the internal state of the output trellis. Thus, the decoding process of $\Y_{\alpha}(z)$ and $\Y_{\beta}(z)$ are shown in Table \ref{table3} and \ref{table4}, respectively.

In Table \ref{table3}, we can clearly see that two conflictions occur in window $[4,6]$ and $[5,7]$, respectively. In window $[4,6]$, combined error vectors $(2\ 1\ 0)(5)$ and $(0\ 0\ 0)(4)$ both will converge to the same trellis output of the same state. So we select $(0\ 0\ 0)(4)$ due to its less weight. The case is same with $(0\ 0\ 0)(4)$ and $(3\ 1\ 0)(5)$ in window $[5,7]$, we eventually select $(0\ 0\ 0)(4)$. At last, by backtracking the father path from $(0\ 0\ 0)(4)$ in window $[6,8]$, we can obtain the MAP path $(1\to0\to1\to0\to0\to1)$ which is identical to $\X(z)$. In fact, $(1\to0\to1\to1\to0\to1)$ is also a possible path because its corresponding $\e(z)=(1+z^5\ 1+z^5\ z^5\ 0\ 0)$ that can yield the same $\Y_{\alpha}(z)$. However its path weight is 5 more than that of the MAP path so that the algorithm discards it. We also can notice the whole decoding process of $\Y_{\alpha}(z)$ must be carried out globally as the algorithm is unable to determine the unqiue MAP path in halfway.

In Table \ref{table4}, we notice that the remaining combined error in window $[0,2]$ and $[3,5]$ are both unique suggesting that network errors are distinguishable in window so that the input sequence can be uniquely determined. Therefore, in window $[0,2]$, the algorithm can promptly output $(1)$ as the first newly found input sequence. Similarly, $(0\to1\to0)$ is output in window $[3,5]$ as the second newly found sequence and $(0),(1),(0)$ are output in window $[4,6],[5,7],[6,8]$, respectively. Eventually, the MAP path $(1\to0\to1\to0\to0\to1)$ is obtained in a distributed way. In fact, all combined errors within $\e_{\beta}(z)$ are within the error capability of $G_{O,t_1}(z)$, which is why a distributed decoding can be realized.
\begin{table}[t]
\centering
\caption{Decoding process of $\Y_{\alpha}(z)$ with $l=2$}
\begin{tabular}{cccccccc}
\toprule
window & [0,2] & [1,3] & [2,4] & [3,5] & [4,6] & [5,7] & [6,8]\\\midrule
\multirowcell{12}{combined\\error\\vectors} & 0 1 1 (0)& \sout{1 1 3} ($\infty$) & 0 0 2 (3) & 0 2 3 (3) & \sout{2 3 0} ($\infty$) & \sout{1 1 3} ($\infty$) & 0 0 0 (4)\\
 & \sout{2 1 2} ($\infty$) & 3 1 0 (3) & 0 2 0 (2) & \sout{2 0 2} ($\infty$) & 0 1 1 (2) & \sout{3 3 2} ($\infty$) & \\
 & 0 3 1 (0) & \sout{1 3 3} ($\infty$) & 0 0 0 (2) & 0 0 1 (2) & 0 3 3 (3) & \uline{0 0 0} (4) & \\
 & \sout{2 3 2} ($\infty$) & 0 0 2 (2) & 0 0 0 (3) & 0 0 3 (3) & \sout{2 1 2} ($\infty$) & 3 1 0 (5) & \\
 & 0 1 3 (0) & \sout{3 3 0} ($\infty$) & 0 2 2 (2) & \sout{2 2 2} ($\infty$) & 2 1 0 (5) & \sout{1 3 1} ($\infty$) & \\
 & 2 1 0 (2) & \sout{1 1 1} ($\infty$) & 0 0 2 (2) & 0 2 1 (2) & \uline{0 0 0} (4) & & \\
 & 0 3 3 (0) & \sout{3 1 2} ($\infty$) &  & 0 2 1 (3) & 0 3 1 (2) &  & \\
 & \sout{2 3 0} ($\infty$) & \sout{1 3 1} ($\infty$) &  & 2 0 0 (4) & 0 1 3 (3) &  & \\
 &  & 0 0 0 (2) &  & 0 0 3 (2) & \sout{2 3 2} ($\infty$) &  & \\
 &  & \sout{3 3 2} ($\infty$) &  & 0 0 1 (3) &  &  & \\
 &  &  &  & \sout{2 2 0} ($\infty$) &  &  & \\
 &  &  &  & 0 2 3 (2) &  &  & \\\bottomrule
\end{tabular}
\label{table3}
\end{table}

\begin{table}[h]
\centering
\caption{Decoding process of $\Y_{\beta}(z)$ with $l=2$}
\begin{tabular}{cccccccc}
\toprule
window & [0,2] & [1,3] & [2,4] & [3,5] & [4,6] & [5,7] & [6,8]\\\midrule
\multirowcell{8}{combined\\error\\vectors} & \sout{1 0 1} ($\infty$) & 0 0 1 (1) & 0 1 0 (1) & \sout{1 0 2} ($\infty$) & 0 0 0 (2) & 0 0 0 (2) & 0 0 0 (2)\\
 & \sout{3 0 2} ($\infty$) & 0 0 3 (1) & 0 3 0 (1) & \sout{3 0 1} ($\infty$) &  &  & \\
 & \sout{1 2 1} ($\infty$) &  & 0 1 2 (1) & \sout{1 2 2} ($\infty$) &  &  & \\
 & \sout{3 2 2} ($\infty$) &  & 0 3 2 (1) & \sout{3 2 1} ($\infty$) &  &  & \\
 & \sout{1 0 3} ($\infty$) &  &  & 1 0 0 (2) &  &  & \\
 & 3 0 0 (1) &  &  & \sout{3 0 3} ($\infty$) &  &  & \\
 & \sout{1 2 3} ($\infty$) &  &  & \sout{1 2 0} ($\infty$) &  &  & \\
 & \sout{3 2 0} ($\infty$) &  &  & \sout{3 2 3} ($\infty$) &  &  & \\\bottomrule
\end{tabular}
\label{table4}
\end{table} 
\end{exam}

\begin{exam}
Assume $G_I(z)=(1+z^2+z^3\ 1+z+z^2+z^3)$ and the network is the same as in Example \ref{exam01}. Thus, the output generators at sink node $t_1,t_2$ are as follows.
\begin{align*}
	&G_{O,t_1}(z)=G_I(z)M_{t_1}(z)=[1+z^2+z^3\ z^2+z^3+z^4]\\
	&G_{O,t_2}(z)=G_I(z)M_{t_2}(z)=[1+z^2+z^3\ 1+z+z^2+z^3]
\end{align*}
Both output generators share the same free distance of $6$. All elements with $\V_0\ne\bm{0}$ in window $[0,6]$ of $G_{O,t_1}(z)$ and in window $[0,3]$ of $G_{O,t_2}(z)$ are given in Table \ref{table5} and Table \ref{table6}. For sink node $t_1$, it can be checked that no addition of arbitrarily two combined error vectors in $\Delta(t_1,l)$ is identical to a certain element in Table \ref{table5} when $l=6$. For sink node $t_2$, the case is same with combined error vectors in $\Delta(t_2,l)$ and elements in Table \ref{table6} when $l=3$. It implies that if all network errors are separated by $7$ timeslots, our decoding algorithm is able to decode messages in a distributed way at both $t_1$ and $t_2$. We can notice that although the free distance of two output generators does not meet the condition in Proposition \ref{prop02}, the distributed decoding can still be realized.
\end{exam}
\begin{table}[t]
\centering
\caption{Elements with $\V_0\ne\bm{0}$ in $\Phi(t_1,l)$ with $l=6$}
\begin{tabular}{cccc}
\toprule
$\X(z)G_{O,t_1}(z)$ in window $[0,l]$ & $\X(z)G_{O,t_1}(z)$ in window $[0,l]$ & $\X(z)G_{O,t_1}(z)$ in window $[0,l]$ & $\X(z)G_{O,t_1}(z)$ in window $[0,l]$\\\midrule
 10 00 11 11 01 00 00 & 10 00 01 11 10 11 01 & 10 10 11 00 10 01 00 & 10 10 01 00 01 10 01\\
 10 00 11 11 01 00 10 & 10 00 01 11 10 11 11 & 10 10 11 00 10 01 10 & 10 10 01 00 01 10 11\\
 10 00 11 11 01 10 00 & 10 00 01 11 10 01 01 & 10 10 11 00 10 11 00 & 10 10 01 00 01 00 01\\
 10 00 11 11 01 10 10 & 10 00 01 11 10 01 11 & 10 10 11 00 10 11 10 & 10 10 01 00 01 00 11\\
 10 00 11 11 11 00 11 & 10 00 01 11 00 11 10 & 10 10 11 00 00 01 11 & 10 10 01 00 11 10 10\\
 10 00 11 11 11 00 01 & 10 00 01 11 00 11 00 & 10 10 11 00 00 01 01 & 10 10 01 00 11 10 00\\
 10 00 11 11 11 10 11 & 10 00 01 11 00 01 10 & 10 10 11 00 00 11 11 & 10 10 01 00 11 00 10\\
 10 00 11 11 11 10 01 & 10 00 01 11 00 01 00 & 10 10 11 00 00 11 01 & 10 10 01 00 11 00 00\\
 10 00 11 01 01 11 11 & 10 00 01 01 10 00 10 & 10 10 11 10 10 10 11 & 10 10 01 10 01 01 10\\
 10 00 11 01 01 11 01 & 10 00 01 01 10 00 00 & 10 10 11 10 10 10 01 & 10 10 01 10 01 01 00\\
 10 00 11 01 01 01 11 & 10 00 01 01 10 10 10 & 10 10 11 10 10 00 11 & 10 10 01 10 01 11 10\\
 10 00 11 01 01 01 01 & 10 00 01 01 10 10 00 & 10 10 11 10 10 00 01 & 10 10 01 10 01 11 00\\
 10 00 11 01 11 11 00 & 10 00 01 01 00 00 01 & 10 10 11 10 00 10 00 & 10 10 01 10 11 01 01\\
 10 00 11 01 11 11 10 & 10 00 01 01 00 00 11 & 10 10 11 10 00 10 10 & 10 10 01 10 11 01 11\\
 10 00 11 01 11 01 00 & 10 00 01 01 00 10 01 & 10 10 11 10 00 00 00 & 10 10 01 10 11 11 01\\
 10 00 11 01 11 01 10 & 10 00 01 01 00 10 11 & 10 10 11 10 00 00 10 & 10 10 01 10 11 11 11\\\bottomrule
\end{tabular}
\label{table5}
\end{table}
\begin{table}[t]
\centering
\caption{Elements with $\V_0\ne\bm{0}$ in $\Phi(t_2,l)$ with $l=3$}
\begin{tabular}{cc}
\toprule
$\X(z)G_{O,t_2}(z)$ in window $[0,l]$ & $\X(z)G_{O,t_2}(z)$ in window $[0,l]$\\\midrule
11 01 11 11 & 11 10 10 00\\
11 01 11 00 & 11 10 10 11\\
11 01 00 10 & 11 10 01 01\\
11 01 00 01 & 11 10 01 10\\\bottomrule
\end{tabular}
\label{table6}
\end{table}

\section{Simulation Results of the Decoding Algorithm}\label{secII}
\subsection{A probabilistic error model}
We define a probabilistic error model for a single source network $G(V,E)$ by defining the probabilities of any set of $i\ (1\le i\le|E|)$ edges of the network being in error at any given time instant as follows. Across time instants, assume that the network errors are subject to i.i.d. according to this distribution.
\begin{align}
	&Pr(i\text{ network edges being in error})=p^i\label{pr01}\\ 
	&Pr(\text{no edges are in error})=q\label{pr02}
\end{align}
where $1\le i\le |E|$ and $0\le p,q\le1$ are real numbers indicating the probability of any single network error in the network and probability of no network error, respectively, such that $q+\sum_{i=1}^{|E|}p^i=1$.  

\subsection{Simulations on the butterfly network}
\begin{figure}[t]
\centering
\includegraphics[scale=1]{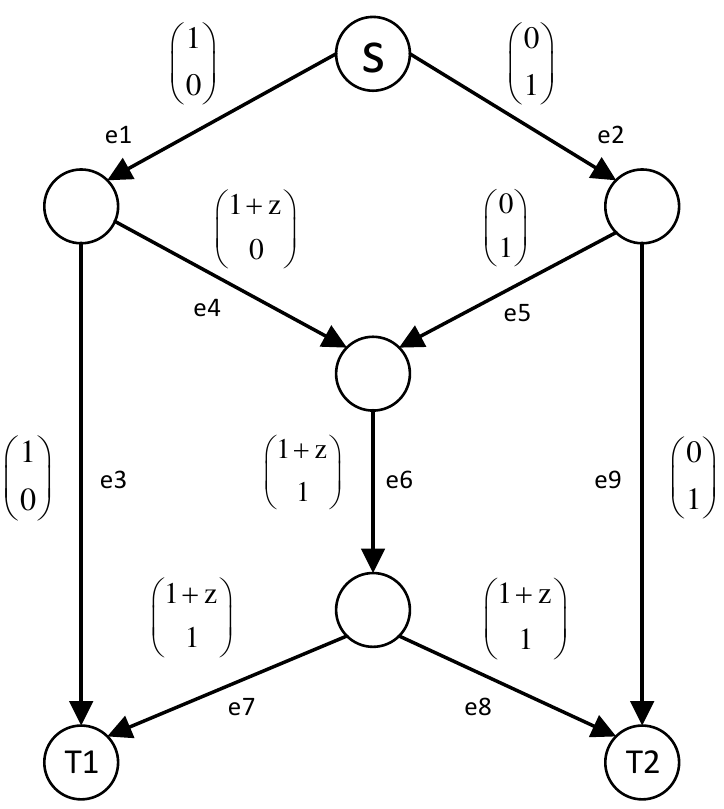}
\caption{Butterfly network $G_2$, with $\omega=2$}
\label{fig2}
\end{figure}

We simulate the proposed algorithm in the classical butterfly network shown in Fig. \ref{fig2}. The transfer matrices for sink node $t_1,t_2$ are given as follows.
\[
	M_{t_1}(z)=\begin{bmatrix}1&1+z\\0&1 \end{bmatrix}=AF_{t_1}(z)
\]
where
\begin{align*}
	A&=\begin{bmatrix}1 & 0 & 0 & 0 & 0 & 0 & 0 & 0 & 0\\0 & 1 & 0 & 0 & 0 & 0 & 0 & 0 & 0 \end{bmatrix}\\
	F_{t_1}(z)&=\begin{bmatrix}1 & 0 & 1 & 0 & 0 & 0 & 0 & 0 & 0\\1+z & 1 & 0 & 1 & 1 & 1 & 1 & 0 & 0\end{bmatrix}^\T
\end{align*}
and
\[
	M_{t_2}(z)=\begin{bmatrix}1+z&0\\1&1 \end{bmatrix}=AF_{t_2}(z)
\]
where
\[
	F_{t_2}(z)=\begin{bmatrix}1+z & 1 & 0 & 1 & 1 & 1 & 0 & 1 & 0\\0 & 1 & 0 & 0 & 0 & 0 & 0 & 0 & 1\end{bmatrix}^\T
\]

Let $l_t=\deg(F_t(z))$. The subspaces of $\Delta(t,l_t)$ for sink node $t_1$ and $t_2$ are shown as in Table \ref{table07} and \ref{table08}, respectively. Thus, all $\Delta(t,l)$ with $l\ge l_t$ can be derived from $\Delta(t,l_t)$ by appending $\bm{0}$ to each combined error vector in the table.
\begin{table}[h]
\centering
\caption{The subspace of $\Delta(t_1,l_{t_1})$ with $l_{t_1}$=1}
\begin{tabular}{cccc}
\toprule
index & minimum error vector $\e$ & $\e F_0\ \e F_1$ & weight \\\midrule
0 & (000000000) & 00 00 & 0\\\midrule
1 & (100000000) & 11 01 & 1\\\midrule
2 & (001000000) & 10 00 & 1\\\midrule
3 & (010000000) (000100000) (000010000) (000001000) (000000100)& 01 00 & 1\\\midrule
4 & (101000000) & 01 01 & 2\\\midrule
5 & (110000000) (100100000) (100010000) (100001000) (100000100)& 10 01 & 2\\\midrule
6 & (011000000) (001100000) (001010000) (001001000) (001000100)& 11 00 & 2\\\midrule
7 & (111000000) (101100000) (101010000) (101001000) (101000100)& 00 01 & 3\\\midrule
\end{tabular}
\label{table07}
\end{table}

\begin{table}[h]
\centering
\caption{The subspace of $\Delta(t_2,l_{t_2})$ with $l_{t_2}=1$}
\begin{tabular}{cccc}
\toprule
index & minimum error vector $\e$ & $\e F_0\ \e F_1$ & weight \\\midrule
0 & (000000000) & 00 00 & 0\\\midrule
1 & (100000000) & 10 10 & 1\\\midrule
2 & (010000000) & 11 00 & 1\\\midrule
3 & (000000001) & 01 00 & 1\\\midrule
4 & (000100000) (000010000) (000001000) (000000010)& 10 00 & 1 \\\midrule
5 & (110000000) & 01 10 & 2\\\midrule
6 & (100000001) & 11 10 & 2\\\midrule
7 & (100100000) (100010000) (100001000) (100000010)& 00 10 & 2 \\\midrule
\end{tabular}
\label{table08}
\end{table}

\begin{figure}
\centering
\includegraphics[scale=1]{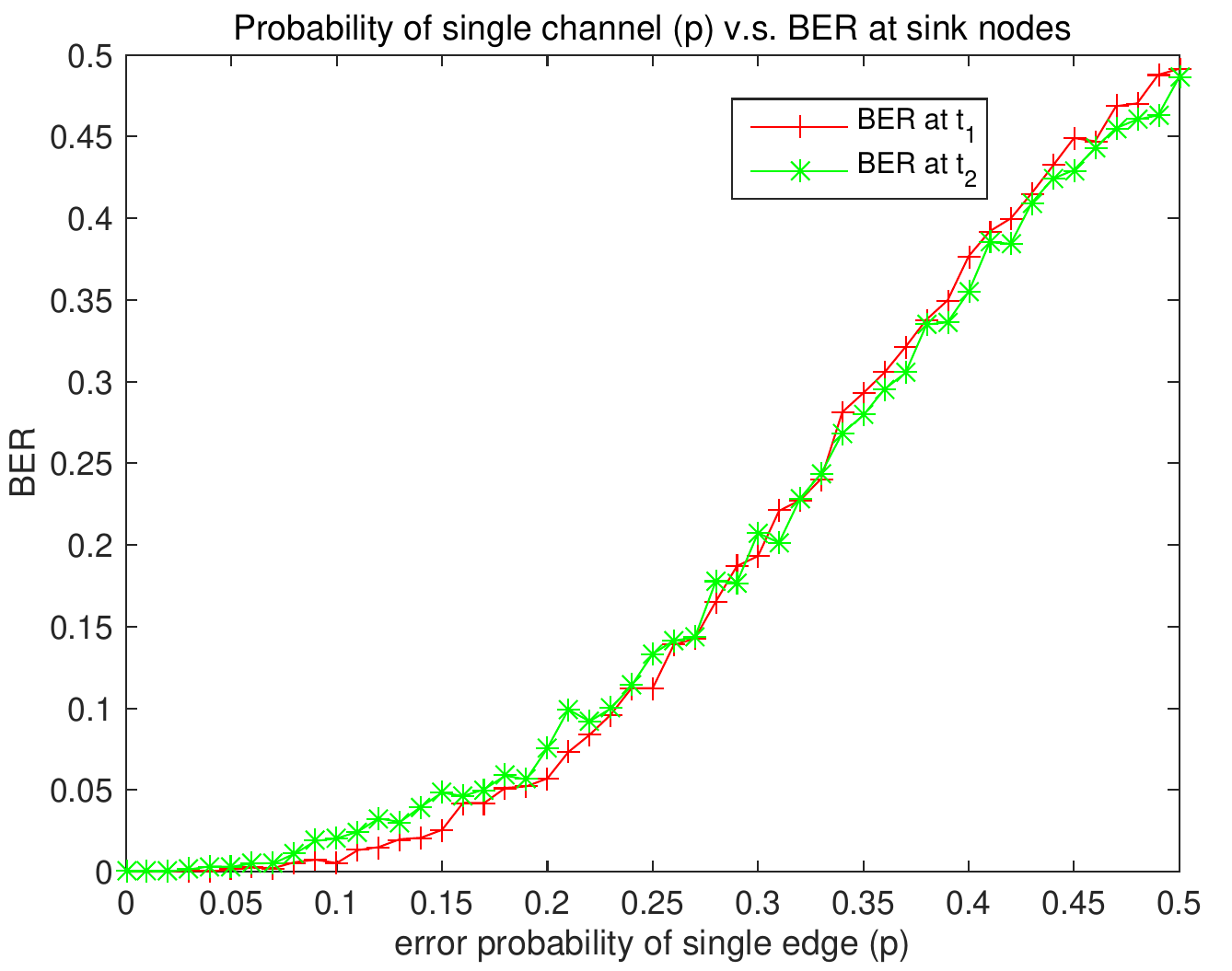}
\caption{BER at both sink nodes}
\label{fig03}
\end{figure}

Let $G_I(z)=[1+z^2+z^3+z^4\ 1+z+z^4]$. The corresponding output generator matrices at sink node $t_1$ and $t_2$ are given as follows.
\begin{align*}
	&G_{O,t_1}(z)=G_I(z)M_{t_1}(z)=[1+z^2+z^3+z^4\ z^2+z^4+z^5]\\
	&G_{O,t_2}(z)=G_I(z)M_{t_2}(z)=[z^2+z^4+z^5\ 1+z+z^4]
\end{align*}
It can be easily checked that $l=2$ satisfying the basic condition that $\Phi(t,l)\cap\Delta(t,l)=\{\zero\}$ for both sink nodes. Let the input sequences be randomly given and error vectors be generated continuously in time under the probabilistic error model (but the decoding algorithm assumes they have been seperated by a certain interval $l$), we compare the decoded sequence obtained by the proposed decoding algorithm with the original input sequence and then compute the average BER given the single edge error probability $p$. The simulation result is shown in Fig. \ref{fig03}.

In Fig. \ref{fig03}, it can be easily observed that the BER performances of our decoding algorithm at sink node $t_1$ and $t_2$ are likely. They both have the following features. In the region where $0\le p\le 0.16$, due to the small single edge error probability, network errors are separated by a sufficient timeslots on the output trellis with a high probability. Therefore the condition of distributed decoding of CNECC is easily satisfied resulting in a better BER performance. In the region where $0.16<p\le0.5$, the network errors on output trellis get closer as the single edge error probability increases. Therefore the condition of distributed decoding of CNECC is no longer satisfied which results in a worse BER performance. However, compared to the simulation results in \cite{004}, we obtain relatively the same BER performance without increasing the decoding complexity by directly decoding messages at sink nodes under the criterion of minimum error weight.

\section{Conclusion}\label{sec5}
In this paper, a Viterbi-like decoding algorithm based on minimum error weight for generalized convolutional coding is proposed and a sufficient condition of realizing distributed decoding of CNECC is first given. The distributed decoding of CNECC is the ideal scenario we strive to pursue. However, the following issues regarding this process still remain open and will be studied in future work. First, how to design a network encoding matrix that is capable of realizing the distributed decoding with a smaller window length. Second, how to select a suitable input generator matrix such that all output generator matrices of all sink nodes are non-catastrophic. Third, how to find a sufficient condition less restricted by $d_{free}$ to realize the distributed decoding of CNECC.

\bibliography{IEEEabrv,references}
\end{document}